\documentclass{IEEEcsmag}

\usepackage[colorlinks,urlcolor=blue,linkcolor=blue,citecolor=blue]{hyperref}
\expandafter\def\expandafter\UrlBreaks\expandafter{\UrlBreaks\do\/\do\*\do\-\do\~\do\'\do\"\do\-}
\usepackage{upmath,color}
\usepackage{subcaption}
\usepackage{xcolor}

\usepackage{etoolbox}
\newtoggle{inclIEEECopyRight}
\toggletrue{inclIEEECopyRight}
\jvol{XX}
\jnum{XX}
\paper{8}
\jmonth{January}
\jname{Computing in Science \& Engineering}
\jtitle{A Unifying Framework to Enable Artificial Intelligence in High Performance Computing Workflows}
\pubyear{2025}

\begin{document}
\iftoggle{inclIEEECopyRight}{
    \begin{titlepage}
    \mbox{}\\{\Large \textbf{IEEE Copyright Notice}}
    \newline\newline\newline\newline
    \textcopyright~2025 IEEE. Personal use of this material is permitted. Permission from IEEE must be obtained for all other uses, in any current or future media, including reprinting/republishing this material for advertising or promotional purposes, creating new collective works, for resale or redistribution to servers or lists, or reuse of any copyrighted component of this work in other works.
    \newline\newline\newline\newline
    {\large Accepted to be Published in: Computing in Science \& Engineering (CiSE) using Digital Object Identifier \href{https://doi.org/10.1109/MCSE.2025.3543940}{10.1109/MCSE.2025.3543940}}
    \end{titlepage}          
}{}

\sptitle{DEPARTMENT: SCIENTIFIC PROGRAMMING}

\title{A Unifying Framework to Enable Artificial Intelligence in High Performance Computing Workflows}

\author{Jens Domke and Mohamed Wahib}
\affil{\mbox{}RIKEN Center for Computational Science, Kobe, 650-0047, Japan}
% Jens Domke: 0000-0002-5343-414X
% Mohamed Wahib: 0000-0002-7165-2095

\author{Anshu Dubey}
\affil{\mbox{}Argonne National Laboratory, Lemont, IL, 60439, USA}
% Anshu Dubey: 0000-0003-3299-7426

\author{Tal Ben-Nun and Erik W. Draeger}
\affil{\mbox{}Lawrence Livermore National Laboratory, Livermore, CA, 94550, USA}
% Tal Ben-Nun: 0000-0002-3657-6568
% Erik W. Draeger: 0000-0003-4063-0253

\markboth{THEME/FEATURE/DEPARTMENT}{THEME/FEATURE/DEPARTMENT}

%template
% https://template-selector.ieee.org/secure/templateSelector/downloadTemplate?publicationTypeId=2&titleId=166&articleId=12&fileId=322
%
%https://www.computer.org/csdl/magazine/cs/write-for-us/14597?title=Author%20Information&periodical=Computing%20in%20Science%20%26%20Engineering
%Department Articles
%Department articles are shorter, between 3,000 and 4,000 words (maximum), including abstract, references, author biographies, and tables/figures, which count as 250 words each. These are only reviewed by the department editors. To pitch or submit a department article, please contact the editor(s) directly using the email address(es) listed on the Editorial Board page.
%
% select a 'department'
%https://www.computer.org/csdl/magazine/cs/about/14589

\begin{abstract}\looseness-1
Current trends point to a future where large-scale scientific applications
are tightly-coupled HPC/AI hybrids. Hence, we
urgently need to invest in creating a seamless, scalable framework where HPC and
AI/ML can efficiently work together and adapt to novel hardware and vendor libraries without
starting from scratch every few years. The current ecosystem and sparsely-connected
community are not sufficient to tackle these challenges, and we require
a breakthrough catalyst for science similar to what PyTorch enabled for AI.
\end{abstract}

\maketitle

\chapteri{T}he potential for scientific discovery is on the cusp of a seismic
change because of a confluence of factors, such as rapid advances in
Artificial Intelligence and Machine Learning (AI/ML) and their demand for low-precision, diverse and 
customizable hardware, and rapid advances in traditional High
Performance Computing (HPC). It is imperative for the scientific
community to start building sustainable software infrastructure that can harness
this potential effectively. Challenges abound because of multiple axes
of growth and their need to keep up and interoperate with one
another. To prepare for the arrival of exascale platforms, the United
States Department of Energy launched the Exascale Computing
Project~(ECP)\footnote{\url{https://www.exascaleproject.org}} seven years ahead of the delivery of the first machine,
Frontier, at Oak Ridge National Laboratory. This endeavor was a
concerted effort to consolidate the gains of two decades
of research and growth in HPC software and hardware and computational science into a robust computing ecosystem.

It would be impossible to repeat the ECP for every generation of new development in hardware,
especially because it is becoming harder to predict what that hardware
might look like. Additionally, the explosive growth in
AI/ML utility, scaling, and national interest\footnote{\url{https://www.youtube.com/watch?v=NFwZi94S8qc}}
occurred during the lifetime of the ECP, but only a handful of
mission-critical codes were in a position to benefit from it.
In part, this is linked to the fact that commercial competition in the AI
space led to rapid developments of siloed software ecosystems which lack
uniform, stable, and scalable APIs to interface with conventional HPC software.
%Furthermore, AI/ML has so much
%commercial competition that the developments are often siloed and the 
%scalability required to interface with conventional HPC does not
%exist.

Together, these developments have created a perfect storm where
the software ecosystem to exploit the advances in HPC and AI/ML
together for science is nearly nonexistent.
Computational science needs immediate investment and commitment to
develop a framework where HPC and AI/ML interoperate efficiently, scale
well, and can evolve with the challenges of growing complexity in
workflows without having to go back the drawing board for each new
generation of platforms, models, and algorithms.

Similar to siloed AI software ecosystems~\cite{riken__2024}, a siloed funding and development
structure fostered the state of HPC software that we are in today, which on the surface appears
coherent. However, if one would create a compatibility matrix---not just
theoretically but on a practical level---of all packages in Spack\footnote{\url{https://spack.io}},
the result would be much sparser than expected. Incentive structures paired
with these incompatibilities in solvers, libraries, languages, tools, etc., then
leads to further inefficiencies, fragmentation and redundant research and developments.
While challenging the status quo is occasionally beneficial~\cite{rajovic_low_2013},
a consolidation of efforts, e.g., as seen recently in the compiler~\cite{lattner_llvm_2004}
or the performance analysis communities~\cite{knupfer_score-p_2012}, benefits everyone
but especially the ones who matter most: the users!
%\newpage

Another dismal example is the various programming frameworks, especially when
targeting accelerators. Although successful in their intended goal, state-of-the-practice
approaches to programming frameworks lack a holistic program view that is
crucial for attaining performance, approachability, and interoperability.
Frameworks often focus on executing task graphs and providing portable
computations at the kernel level, which are only a part of modern-day performance
engineering. Scientific codes, however, are workflows consisting of multiple
components, i.e., languages, numerical and communication libraries, frameworks,
surrogate AI models, and runtime memory management---all of which need to be
carefully orchestrated to minimize memory footprint and data movement.
As a result of this decoherence, many optimization opportunities are missed.
%\noter{Jens, somewhere in here it would be good to have your observations about
%how various attempts by the tool developers have failed and why we need to take a different approach}

We are a group of computer and computational scientists who believe that 
%have had
%deep involvement in several application projects and in procurement of
%leadership class systems, and therefore, have extensive
%first hand experience of challenges faced by the science teams. 
the HPC community should be deeply concerned about the imminent barrier to
innovation that will come from not being able to use future resources in a
timely fashion. Additionally, the next iteration of code refactoring must
take a fundamentally different approach than the ECP, because the pace of
change in hardware and its specialization is likely to be faster than the
pace in which a typical team can adapt their code(s) to the target hardware.
%We urgently need to start planning for 
Therefore, we have put together ideas---based on our own prior experiences---for
a viable application-facing framework, that, we hope and expect
would be able to support a large class of tightly-coupled HPC/AI hybrids through
several generations of hardware evolution.
%in interacting with complex scientific workflows. 
%a proposal for development of
%an HPC-AI/ML computational science framework to meet science needs. 
%a wide variety of (even perhaps all) 
Here, we describe our ideas 
%for such a framework 
that we believe 
%We believe that we have experience, knowledge, and motivation to put
%together an international effort to build a flexible and extensible
%framework that 
can not only meet the needs of several mission-critical science domains, but also provide
a blueprint for software where our design may not be directly applicable.
%directly served by our proposed framework.
%\vadjust{\pagebreak} 
\vspace*{-5pt}

\section{LATEST IN SCIENTIFIC PROGRAMMING}
With the exception of the C++\,-based tool
%s such as 
Kokkos\footnote{\url{https://kokkos.org}}, which has seen some adoption, 
%, Raja, and GridTools,
the vast majority of abstractions, programming models, and performance
portability tools have not been adopted by more than a handful of
scientific applications. Kokkos effectively addresses the challenge of 
having to maintain multiple code variants for every different hardware
target. As a result, cottage industries have grown in many parts of
the world where codes are being  converted from Fortran to C++, only
to be able to use Kokkos. Those efforts still focus on a largely
dominant parallel execution model---it has merely switched from
distributed memory to a massively parallel model.

For many of the
codes in the ECP, the objective was to offload as much work as possible
to the GPU, because of its computational and energy efficiency.
Orchestration of data and computation movement received very limited
attention, and even more limited adoption. Orchestration was meant
to be the purview of 
%was expected to be the purview of
the task-based runtime systems. Although several have been under
development for years,
%~\cite{duran_ompss_2011}, 
the only success stories they have are when
the applications using them were co-designed with the tool itself.
Legacy codes, or independently developed codes have had no success
with them without deep refactoring, or in a few cases, completely
rewriting the code. The latter was up to now unfeasible in our community
%~\cite{sorensen_special_2019},
and whether AI-assisted code translation will be possible
%~\cite{dearing_lassi_2024} 
without sufficient
training data remains an open research question.

Another rapidly growing trend is that Python now dominates as the language of choice for algorithmic innovations in many scientific domains.
This trend has been visible for a few years, due to a rich supporting
ecosystem and the ease of building prototypes with it. The arrival
of AI/ML frameworks with Python interfaces has further accelerated
this trend across industry and academia. Transitioning prototypes developed in Python---especially
those that are in flux---to performant HPC codes has emerged as a
real challenge. We are aware of at least one use case where their
toolchain\footnote{\url{https://gridtools.github.io/gt4py/latest/gtscript.html}}
pivoted from being C++\,-centric to Python-centric because
of domain scientists' discomfort with C++, thereby
making the use of C++\,-based abstractions an unsustainable development model. 
%interfaces to python
%interfaces with a large group dedicated to development of intermediate
%stages of translation for a single class of %applications~\cite{FIXME}.

It may seem impossible to reconcile the growth of Python-based
development with the anticipated growth in hardware
complexity. However, the path taken by the AI/ML frameworks---enabling
plug-and-play code modules with some degree of customization---hints
at a way forward for HPC codes as well. For a scientist, a dream come true
would be to have an ecosystem where many capabilities they need
already exist, and where they can plug in missing capabilities to get
results at the cost of having to conform to the coding standards of
the framework. Such an approach had some notable successes in the cluster computing era with multiphysics code frameworks that enabled plug-and-play features for physics solvers\footnote{E.g. \url{https://www.cactuscode.org}, ~\url{https://flash-x.org}, ~\url{https://www.lammps.org}, and ~\url{https://www.gromacs.org}.}. 
%~\cite{goodale_cactus_2003, dhruv_framework_2023, thompson_lammps_2022, pall_heterogeneous_2020}.
We have succeeded in
achieving some of these objectives in narrower contexts~\cite{DubeyWeideONeilEtAl22,Ben-Num2022}.

We believe that our collective experience positions us well to %design an 
%ecosystem 
articulate the design challenges and possible solutions that may meet the
needs of domain scientists who require large-scale supercomputers to tackle
mission-critical and societal concerns.\vspace*{-5pt}

\begin{figure*}[ht!]
    \centering
    \begin{minipage}{0.52\textwidth}
    \centerline{\includegraphics[height=1.85in,clip,trim={0.5in 1.5in 0.5in 1.5in}]{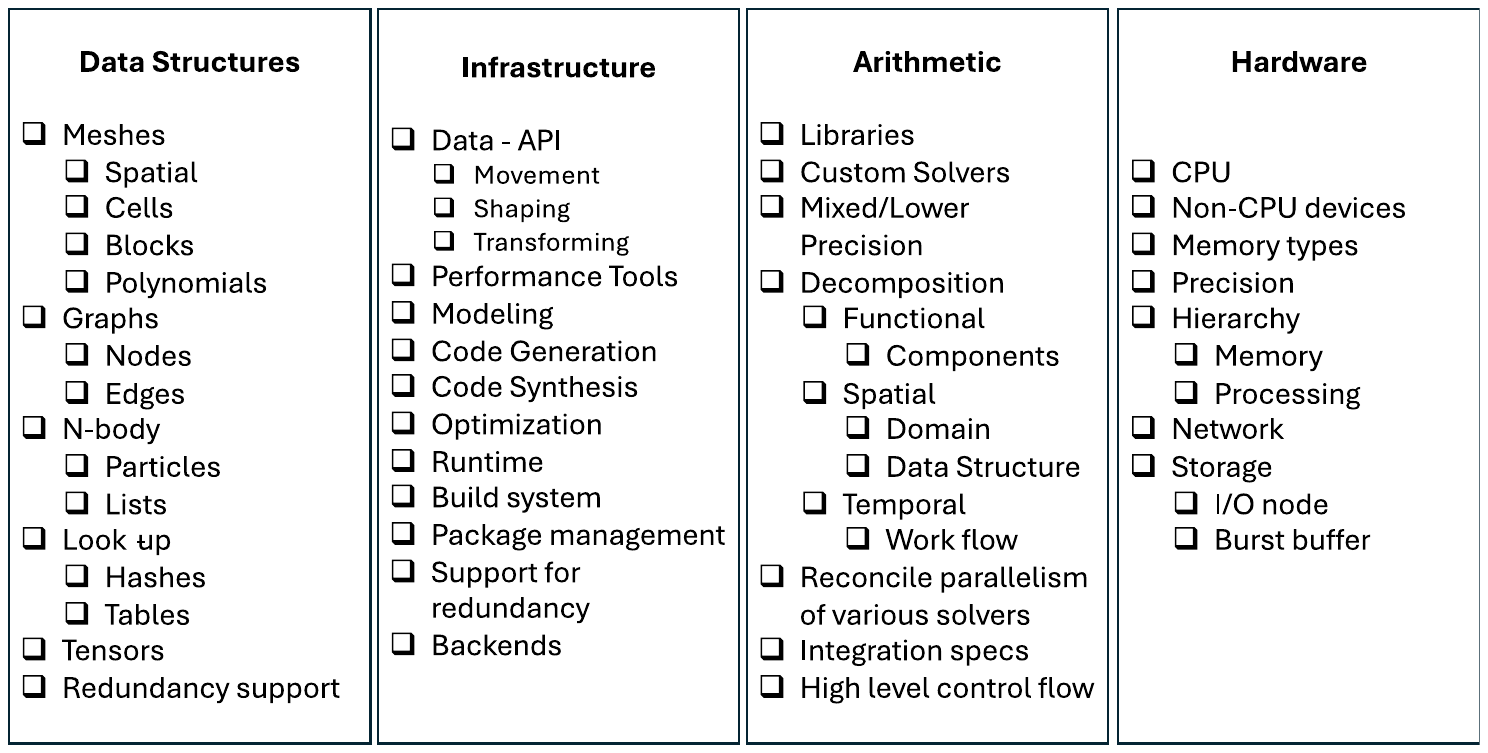}}
    \caption{The design space that a multiphysics HPC framework must take into account already in the absence of AI/ML integration or offloading.}
    \label{fig:app-needs}\vspace*{-0pt}
    \end{minipage}
    \hfil
    \begin{minipage}{0.09\textwidth}
    \centerline{\includegraphics[height=0.5in,clip]{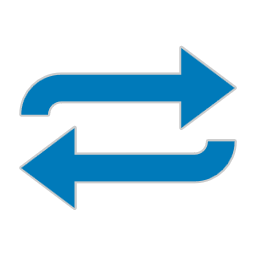}}
    \end{minipage}\hspace{-25pt}
    \begin{minipage}{0.38\textwidth}
    \centerline{\includegraphics[height=1.85in,clip,trim={0.5in 1.5in 5.5in 1.5in}]{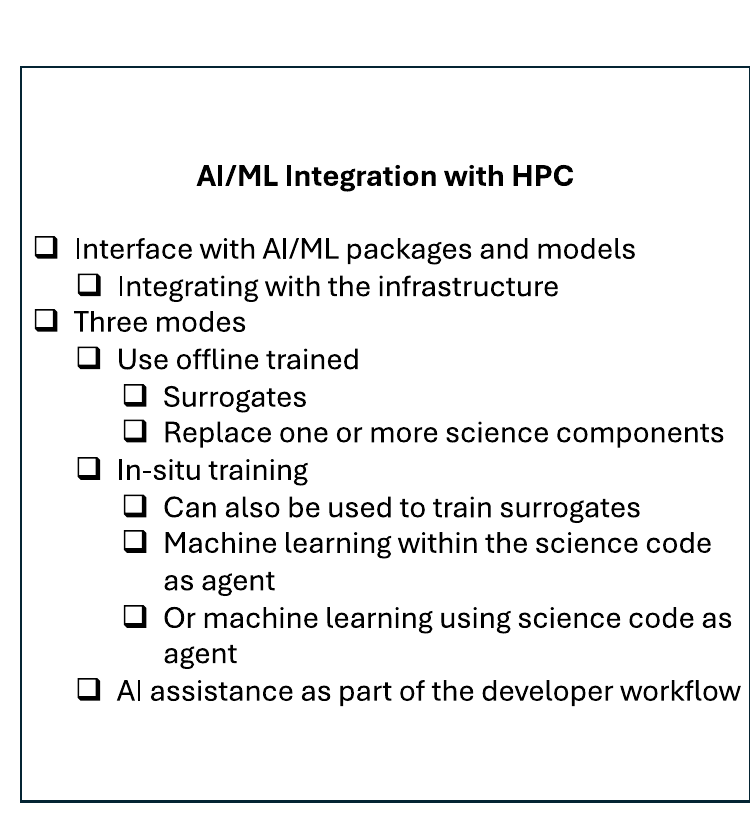}}
    \caption{Various ways in which we anticipate use of AI/ML in High-Performance Computing software.}
    \label{fig:app-and-ai}\vspace*{-0pt}
    \end{minipage}
\end{figure*}

\section{OBJECTIVES AND FEATURES FOR THE FRAMEWORK}
We are aware that this is an extremely complicated design space, and that any attempt to implement a general purpose solution would be a very ambitious undertaking. However, we also believe that a concerted effort to crystallize some ideas for how such a framework can be built is already past due. If we had a design and started building the framework now we may have had a possibility of having a partially functional one halfway through the life of the post-exascale platforms.  Additionally, past experiences show that attempting to fully specify the design of all aspects of any framework without substantial input from the end-users, the scientists, do not survive the test of deployability. Therefore, the exploration of the design space could, in itself, be a multi-year effort. 

The only viable approach at this late date is for the design specifications to be living documents, which may be subject to change throughout the life of the framework. Meaningful in-depth input from end-users must be central to the evolution of the design. Certain insights from existing frameworks and abstraction tools point to features, listed in Figure~\ref{fig:app-needs}, which will definitely be needed in the framework. They include a library of data structures for the supported methods and models compatible with known algorithmic constraints; tools to efficiently transform data when migrating between supported data structures and hardware; and embedded tools to monitor and report on runtime behavior. The latter will be necessary for debugging, reasoning about performance, as well as training the AI/ML engines. Figure~\ref{fig:app-and-ai} shows our current view of how
AI/ML will be used in this ecosystem. Effectively serving a wide range of science areas, algorithms, and implementation strategies hinges on a set of features, listed hereafter, which are essential to building and maintaining scalable HPC-AI/ML applications.

\subsection{\textit{Desired Features:}}
%\looseness-1All acronyms ...
Below we enumerate the high-level features that we believe the framework must have to be successful. 
\begin{itemize}
\item[{\ieeeguilsinglright}] {\bf Data transformers}. Components that can efficiently
convert data structures for different portions of the workflow, i.e., from an
HPC friendly data container to AI/ML friendly one. 
\item[{\ieeeguilsinglright}] {\bf Disentangling math from control flow}. The
holy grail for domain scientists is to write their equations and have the
code generated to solve them. While that is impossible to achieve universally
given the multitudes of ways in which  equations can be solved, it is
possible to enable implementation of numerical algorithms without
entanglement with data structures through appropriate abstractions.
\item[{\ieeeguilsinglright}] {\bf Expressing locality}. Being able to
explicitly state the data and computation locality is a critical
necessity for efficiency and energy saving.
\item [{\ieeeguilsinglright}] {\bf Constraining semantics}. Idiomatic richness of the target language is a proportionate burden on the optimizing compiler. Equations do not need this richness; thus, a semantically constrained language subset that is mutually agreed upon by the domain developers and tool developers benefits both sides, and ultimately science. Domain-specific languages follow this idea but relying on new syntax makes them niche solutions.
\item[{\ieeeguilsinglright}] {\bf Composability}. A very high degree
of composability is essential for any framework that aims to
support a wide range of scientific workflows while also accommodating non-public code.
\item[{\ieeeguilsinglright}] {\bf Cost models}. Such models assist in composing
the components of an application and guide optimal workflow execution decisions. 
\end{itemize}
The lower levels---especially where competing solutions exist---will require input and careful selections by panels of subject matter experts and users.
\vspace*{-5pt}

\begin{figure*}
    \centering
    \begin{subfigure}[b]{0.49\textwidth}
        \centering
        \includegraphics[width=17.5pc,page=1,clip,trim={0in 1in 0in 0in}]{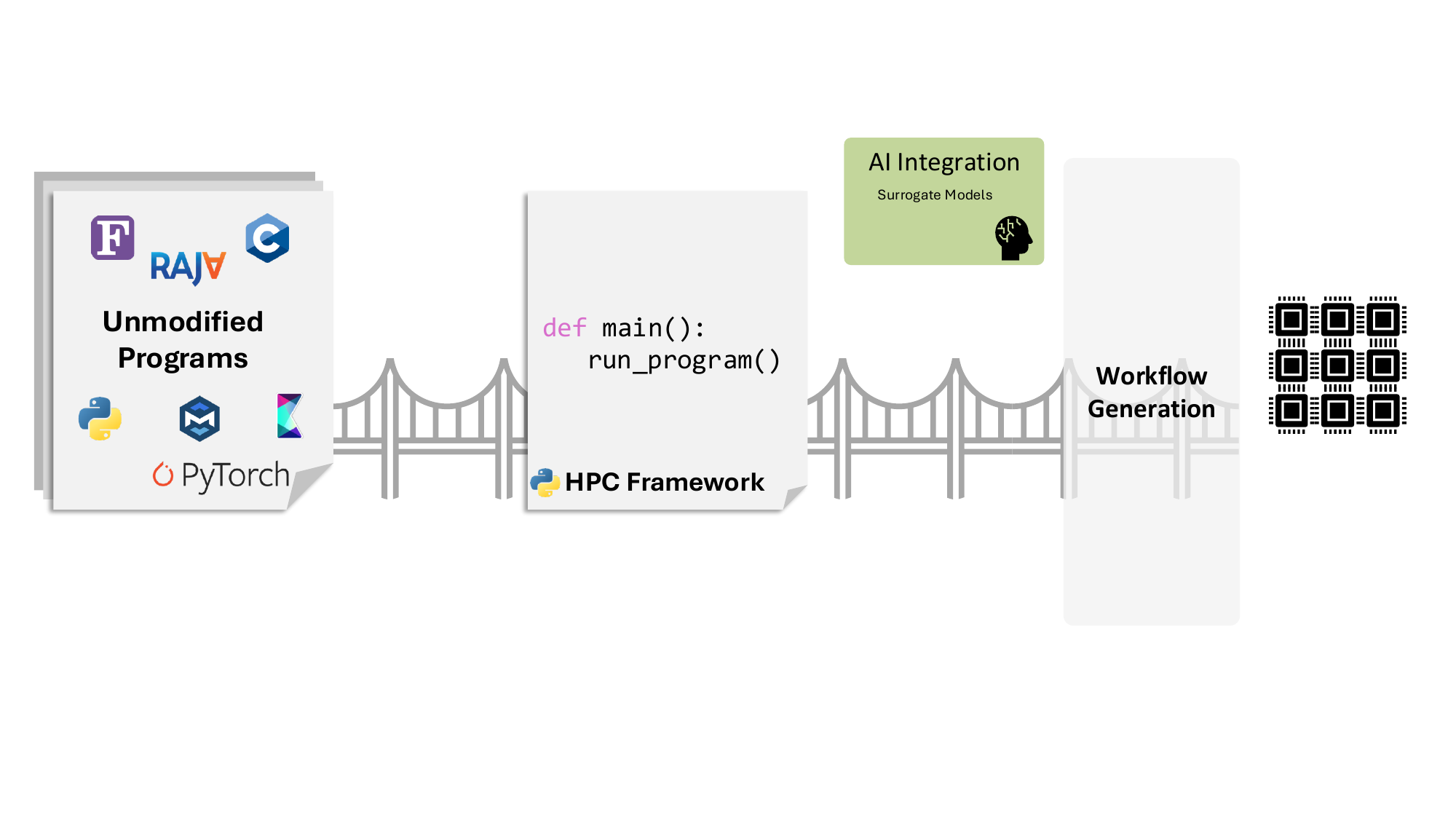}
        \caption{\textbf{Phase 1}: Thin Application Wrapper}
    \end{subfigure}\hfill
    \begin{subfigure}[b]{0.49\textwidth}
        \centering
        \includegraphics[width=17.5pc,page=2,clip,trim={0in 1in 0in 0in}]{Arxiv/figures/phases}
        \caption{\textbf{Phase 2}: Sub-Program Integration}
    \end{subfigure}%
    \vspace{1em}
    \begin{subfigure}[b]{0.49\textwidth}
        \centering
        \includegraphics[width=17.5pc,page=3,clip,trim={0in 1in 0in 0in}]{Arxiv/figures/phases}
        \caption{\textbf{Phase 3}: Retargetable Building Blocks}
    \end{subfigure}\hfill
    \begin{subfigure}[b]{0.49\textwidth}
        \centering
        \includegraphics[width=17.5pc,page=4,clip,trim={0in 1in 0in 0in}]{Arxiv/figures/phases}
        \caption{\textbf{Phase 4}: Algorithmic Description}
    \end{subfigure}%
    \vspace{.8em}
    \caption{Workflow of the proposed HPC framework. Phases enable gradual application integration --- the more information provided, the more capabilities can be unlocked to reduce the overhead of utilizing modern hardware.}
    \label{fig:workflow}
\end{figure*} 

\section{DEVELOPMENT APPROACH}
Our overarching vision is to allow codes to follow a staggered development
and migration. Our collective experience is that any abrupt shift
in software architecture will not only be disruptive for science that is
relying upon the software for discovery, but also a tremendous
barrier of entry to adopting new approaches. Application codes are large and
complex, and making them compatible with any new programming model or
abstraction takes a long time. If no mechanism is provided to keep
the code usable while the shift is taking place, scientists cannot continue
their work.

We propose beginning with a non-intrusive framework that enables bridging of models through
interfaces without any attempts to add new abstractions. This should allow
applications to plug into the framework and be able to use features that
they may not already have, as shown in Phase 1 of
Figure~\ref{fig:workflow}. The applications may be able to use limited
AI/ML interfaces. As we progress towards later phases, we begin to
introduce abstractions where, in exchange for handing some of the
control and details over to the  framework, the applications gain greater
performance and portability while simplifying their own code base.
For example, in Phase 2 the framework may become the primary custodian
of data containers, with applications indicating what their needs
are, allowing them to hand over the communication and data movement to the
framework. In the next phase, we can further abstract the knowledge of
hardware and control of data from the application so that the
framework can internally apply transformations and optimizations as
needed. Our final goal is to reach a state where the algorithm writer
can express their arithmetic and data storage requirements without
binding them to one another. Instead, they can leave the management to
the framework.

The use of AI/ML elements is interwoven into the various phases, so
as to allow sub-programs to be seamlessly exchanged for surrogate models,
as well as use AI to estimate performance models and facilitate
automatic optimization without brute-force search.

As the framework grows more sophisticated, one can bring in the
state-of-the-art code generation/translation/assembly techniques as
integrated capabilities into the framework. Simultaneously, one can
integrate necessary tooling to visualize and analyze results,
as well as monitor and understand performance.\vspace*{-5pt}

\section{IMPLEMENTATION PROPOSAL}
Given that our HPC community already has a rich ecosystem of existing
libraries and tools, we envision a two-pronged approach to the development
of the framework. One would begin with a pilot where one 
can simultaneously study and digest ways in which the existing abstractions
and tools are hampered from interoperability by their
design choices, and conceptualizing the elements of the framework that
would overcome these limitations. In this phase, the
backbone elements of the framework would be prototyped and evaluated by the
stakeholders and the wider community. This is where the tight coupling with
the end user becomes critical for success.

Because we envision existing
libraries and tools being integrated into the whole with appropriate
refactoring, the approach would be to make a draft design for some
selected (usually critical) portion of the library as a component, build the substrate in the framework where the  component fits, and then refactor the section of the library to fit into the framework. At this point, it would be feasible to partially evaluate the scalability and efficacy of the design. One could assess the gaps and weaknesses and tweak it as needed. In some circumstances some components of the design may even need a hard reset. The same process can be repeated  with other sections of the library,
going back to any of the earlier refactored portions and substrates
and modifying them if necessary.

Through this incremental and cooperative co-design approach, we should be
able to catch design flaws early in the development cycle. We envision several teams working on different
aspects of the framework, who are assisted by the latest large-language models to accelerate
mundane tasks such as documentations, code transpilations, and interface creation.
However, there should always be a degree of overlap among teams and a continuous exchange of information, especially if any team makes non-trivial change in its direction.\vspace*{-5pt}

%\noter{Since we have space, we could give a few more details about the two use-cases we mentioned earlier. What do you think?}

\section{CONCLUDING REMARKS}
We agree that this is no trivial undertaking.
However, the Linux Kernel community, and more recently Deep Learning community
with PyTorch et al., have shown what can be achieved by coordinated and
targeted efforts. Unfortunately, thus far, HPC lacks this level of 
coordination, resulting in the field being left behind in the rapidly changing landscape
of hardware and software options. We simply cannot afford that
every application team struggles with the same challenges in isolation anymore,
especially when it comes to code modernization and AI/ML integration.

For efficiency reasons, we urgently need a single scalable framework,
into which the majority of relevant scientific algorithms are either
integrated or can seamlessly be glued into, to enable HPC-AI/ML hybrids.
Data transformers, reuse and composability, code generation, and separation of concerns are the keys
to allow computational scientists to focus on algorithms, while computer
scientists tackle the efficiency challenges involved in mapping these
algorithms onto various processors and distributed
architectures.\vspace*{-5pt}

%\noter{I have mostly just changed the tone from "we will do it" to "these are our ideas, we think it should be done this way" just to make it read more like an editorial and less like a proposal. Most of the rest of the content is I think in pretty good shape for this kind of article.}

% After appearance
\section{ACKNOWLEDGMENTS}
% The author(s) would like to thank ... %A, B, and C. This work was supported by XYZ under Grant \#\#\#.
% rick/satoshi,DoE/Mext feedback? any DOE funding stuff?
Argonne National Laboratory's work was supported by the U.S. Department of Energy, Office of Science, under contract DE-AC02-06CH11357
and LLNL work was supported under Contract DE-AC52-07NA27344 (LLNL-JRNL-2002577). 
\def\refname{REFERENCES}
\bibliographystyle{IEEEtran}
\bibliography{IEEEabrv,CiSE_refs}\vspace*{-8pt}

\begin{thebibliography}{1}
\providecommand{\url}[1]{#1}
\csname url@samestyle\endcsname
\providecommand{\newblock}{\relax}
\providecommand{\bibinfo}[2]{#2}
\providecommand{\BIBentrySTDinterwordspacing}{\spaceskip=0pt\relax}
\providecommand{\BIBentryALTinterwordstretchfactor}{4}
\providecommand{\BIBentryALTinterwordspacing}{\spaceskip=\fontdimen2\font plus
\BIBentryALTinterwordstretchfactor\fontdimen3\font minus
  \fontdimen4\font\relax}
\providecommand{\BIBforeignlanguage}[2]{{%
\expandafter\ifx\csname l@#1\endcsname\relax
\typeout{** WARNING: IEEEtran.bst: No hyphenation pattern has been}%
\typeout{** loaded for the language `#1'. Using the pattern for}%
\typeout{** the default language instead.}%
\else
\language=\csname l@#1\endcsname
\fi
#2}}
\providecommand{\BIBdecl}{\relax}
\BIBdecl

\bibitem{riken__2024}
{RIKEN}, ``{{Research}} and {{Study}} on {{Next Generation Computing
  Infrastructure}} - {{System Research}} and {{Development}} - {{Result
  Report}} (3),'' RIKEN, Tech. Rep., Jun. 2024,
  \url{https://www.mext.go.jp/content/20241111-mxt-jyohoka01-000038552_03.pdf}.

\bibitem{rajovic_low_2013}
N.~Rajovic, L.~Vilanova, C.~Villavieja, N.~Puzovic, and A.~Ramirez, ``The low
  power architecture approach towards exascale computing,'' \emph{Journal of
  Computational Science}, vol.~4, no.~6, pp. 439--443, 2013.

\bibitem{lattner_llvm_2004}
C.~Lattner and V.~Adve, ``{{LLVM}}: A compilation framework for lifelong
  program analysis \& transformation,'' in \emph{International {{Symposium}} on
  {{Code Generation}} and {{Optimization}}, 2004. {{CGO}} 2004.}, 2004, pp.
  75--86.

\bibitem{knupfer_score-p_2012}
A.~Kn{\"u}pfer, C.~R{\"o}ssel, D.~an~Mey, S.~Biersdorff, K.~Diethelm,
  D.~Eschweiler, M.~Geimer, M.~Gerndt, D.~Lorenz, A.~Malony, W.~E. Nagel,
  Y.~Oleynik, P.~Philippen, P.~Saviankou, D.~Schmidl, S.~Shende,
  R.~Tsch{\"u}ter, M.~Wagner, B.~Wesarg, and F.~Wolf, ``Score-{{P}}: {{A Joint
  Performance Measurement Run-Time Infrastructure}} for {{Periscope}},
  {{Scalasca}}, {{TAU}}, and {{Vampir}},'' in \emph{Tools for {{High
  Performance Computing}} 2011}, H.~Brunst, M.~S. M{\"u}ller, W.~E. Nagel, and
  M.~M. Resch, Eds.\hskip 1em plus 0.5em minus 0.4em\relax Berlin, Heidelberg:
  Springer Berlin Heidelberg, 2012, pp. 79--91.

\bibitem{DubeyWeideONeilEtAl22}
A.~Dubey, K.~Weide, J.~O'Neal, A.~Dhruv, S.~Couch, J.~A. Harris, T.~Klosterman,
  R.~Jain, J.~Rudi, B.~Messer, M.~Pajkos, J.~Carlson, R.~Chu, M.~Wahib,
  S.~Chawdhary, P.~M. Ricker, D.~Lee, K.~Antypas, K.~M. Riley, C.~Daley,
  M.~Ganapathy, F.~X. Timmes, D.~M. Townsley, M.~Vanella, J.~Bachan, P.~M.
  Rich, S.~Kumar, E.~Endeve, W.~R. Hix, A.~Mezzacappa, and T.~Papatheodore,
  ``Flash-{X}: A multiphysics simulation software instrument,''
  \emph{SoftwareX}, vol.~19, p. 101168, 2022.

\bibitem{Ben-Num2022}
T.~Ben-Nun, L.~Groner, F.~Deconinck, T.~Wicky, E.~Davis, J.~Dahm, O.~D. Elbert,
  R.~George, J.~McGibbon, L.~Trümper, E.~Wu, O.~Fuhrer, T.~Schulthess, and
  T.~Hoefler, ``Productive performance engineering for weather and climate
  modeling with {Python},'' in \emph{SC22: International Conference for High
  Performance Computing, Networking, Storage and Analysis}, 2022, pp. 1--14.

\end{thebibliography}

\begin{IEEEbiography} {Jens Domke}
is a Team Leader at RIKEN Center for Computational Science, Kobe, Japan.
His research interests include system co-design, performance evaluation, interconnect
networks, and optimization of parallel applications and architectures.
\end{IEEEbiography}

\begin{IEEEbiography} {Mohamed Wahib}
is a Team Leader at RIKEN Center for Computational Science, Kobe, Japan.
His research interests revolve around the central topic of ``Performance-centric
Software Development'', in the context of HPC and AI.
\end{IEEEbiography}

\begin{IEEEbiography} {Anshu Dubey}
is a Senior Computational Scientist known for her work on multiphysics software
for high-performance computing. She is the chief architect of FLASH and leads
the development of Flash-X. 
\end{IEEEbiography}

\begin{IEEEbiography} {Tal Ben-Nun}
is a Computer Scientist in the Center for Applied Scientific Computing at the
Lawrence Livermore National Laboratory. His research interests involve the
intersections between programming languages, machine learning, and high-performance computing.
\end{IEEEbiography}

\begin{IEEEbiography} {Erik W. Draeger}
is a Group Leader of the Scientific Computing Group in the Center for Applied
Scientific Computing at the Lawrence Livermore National Laboratory.
He co-developed the Cardioid cardiac electrophysiology code, and is one of
the main developers of the HARVEY code that was a 2015 Gordon Bell Award
Finalist.
\end{IEEEbiography}

\end{document}